\documentclass[onecolumn,showpacs,
superscriptaddress,preprintnumbers,amsmath,amssymb]{revtex4}

\usepackage{graphicx}
\usepackage{dcolumn}
\usepackage{bm}

\def\Ren#1{\mbox{\boldmath$#1$}}

\newcommand{\beq}{\begin{eqnarray}}
\newcommand{\eeq}{\end{eqnarray}}

\begin{document}
  
  \preprint{}
  
  \title{
    Second-order Relativistic Hydrodynamic Equations for Viscous Systems;
    how does the dissipation affect the internal energy?
  }
  
  \author{Kyosuke Tsumura}
  \affiliation{
    Analysis Technology Center,
    Fujifilm Corporation,
    Kanagawa 250-0193, Japan
  }
  
  \author{Teiji Kunihiro}
  \affiliation{
    Department of Physics,
    Kyoto University,
    Kyoto 606-8502, Japan
  }
  
  \begin{abstract}
    We derive the second-order dissipative relativistic hydrodynamic equations
    in a generic frame with a continuous parameter
    from the relativistic Boltzmann equation.
    We present explicitly the relaxation terms in the energy and particle frames.
    Our results show that
    the viscosities are frame-independent
    but the relaxation times are generically frame-dependent.
    We confirm that the dissipative part of the energy-momentum tensor
    in the particle frame satisfies
    $\delta T^{\mu}_{\,\,\,\mu} = 0$
    obtained for the first-order equation before,
    in contrast to the Eckart choice
    $u_{\mu} \, \delta T^{\mu\nu} \, u_{\nu} = 0$
    adopted as a matching condition in the literature.
    We emphasize that the new constraint
    $\delta T^{\mu}_{\,\,\,\mu} = 0$
    can be compatible with the phenomenological derivation
    of hydrodynamics based on the second law of thermodynamics.
  \end{abstract}
  
  \date{\today}
  \maketitle
  
  \setcounter{equation}{0}
  \section{Introduction}

  After the discovery that
  perfect hydrodynamics can be valid for describing the phenomenology
  of Relativistic Heavy Ion Collider (RHIC)
  at Brookhaven National Laboratory
  \cite{qcd001, qcd002, qcd003},
  people are now interested in
  relativistic hydrodynamics for \textit{dissipative} systems;
  see the recent excellent review articles
  \cite{Hirano:2008hy,Romatschke:2009im}.

  Recently,
  Tsumura, Kunihiro (the present authors) and Ohnishi (abbreviated as TKO)
  \cite{TKO}
  derived generic covariant hydrodynamic equations
  for a viscous fluid from the relativistic Boltzmann equation
  in a systematic manner with no heuristic arguments
  on the basis of the so-called renormalization group (RG) method
  \cite{rgm001,env001,env004,HK02,env006}.
  Although the hydrodynamic equations they derived are
  the so-called first-order ones,
  the equations have remarkable aspects:
  The generic equation derived by TKO can produce
  a relativistic dissipative hydrodynamic equation
  in any frame with an appropriate choice
  of a macroscopic flow vector $\Ren{a}^{\mu}$
  ($\mu = 0,\, 1,\, 2,\, 3$),
  which defines the coarse-grained space and time;
  the resulting equation in the energy frame
  coincides with that of Landau and Lifshitz
  \cite{hen002},
  while that in the particle frame is similar to,
  but slightly different from, the Eckart equation
  \cite{hen001}.

  Let $\delta T^{\mu\nu}$ and $\delta N ^{\mu}$
  be the dissipative term
  of the symmetric energy-momentum tensor
  and the particle-number vector, respectively.
  Owing to the ambiguity in the separation of the energy and the mass
  inherent in relativistic theories,
  one must choose the
  local rest frame (LRF)
  where the flow velocity
  $u^\mu$ with $u^\mu \, u_\mu = 1$ is
  defined:
  One of the typical frame is the energy (Landau) frame in which
  $\delta T^{\mu\nu} \, u_{\mu} \, \Delta_{\nu\rho} = 0$
  with $\Delta^{\mu\nu} \equiv g^{\mu\nu} - u^\mu \, u^\nu$
  and $g^{\mu\nu} = \mathrm{diag}(+1,\,-1,\,-1,\,-1)$,
  i.e. there is no dissipative energy flow.
  On the other hand, another typical frame is the particle (Eckart)
  frame in which
  $\delta N ^{\mu} \, \Delta_{\mu\nu} = 0$,
  i.e. there is no dissipative particle flow.
  Both in the energy and particle frames,
  the dissipative terms of the energy-momentum tensor
  and the particle-number vector
  are usually assumed to satisfy the constraints,
  \begin{eqnarray}
    \label{eq:PT-E}
    u_{\mu} \, \delta T^{\mu\nu} \, u_{\nu} = 0,
  \end{eqnarray}
  and $u_{\mu} \, \delta N^{\mu} = 0$.
  These phenomenological ansatz have been employed as the matching conditions
  even in the subsequent ``derivations''
  of the so-called second-order equations
  \cite{hen003,mic004,mic001,Betz:2008me};
  note that
  even in the Grad's moment method \cite{grad},
  some ansatz are needed to
  $\delta T^{\mu\nu}$ and $\delta N^{\mu}$ as the matching conditions,
  for which different proposals exist
  \cite{marle:69,mic004}.
  
  Here we emphasize that
  the matching conditions touch on the fundamental but not yet fully understood problem
  how to define the LRF in the relativistic fluid dynamics for a viscous system.
  The way how to define the LRF or equivalently to fix the matching condition is unsolved yet,
  and remains a nontrivial and fundamental problem
  in the field of nonequilibrium relativistic dynamics,
  although there have been no serious consideration on this difficult
  problem in the literature.
  Actually,
  we shall argue that these phenomenological ansatz, especially Eq.(\ref{eq:PT-E}),
  can be false and actually is not compatible with
  the underlying kinetic equation.

  In fact, it is found that
  the TKO equation in the particle frame
  derived from the relativistic Boltzmann equation satisfies
  \begin{eqnarray}
    \label{eq:PT-TKO}
    \delta T^{\mu}_{\,\,\,\mu} = 0,
  \end{eqnarray}
  but does not satisfy Eq.(\ref{eq:PT-E}).
  One should here note that
  the derived condition (\ref{eq:PT-TKO}) is identical to a matching condition
  postulated by Marle
  \cite{marle:69}
  and advocated by Stewart
  \cite{mic003}
  in the derivation of the relativistic hydrodynamics
  from the relativistic Boltzmann equation
  with use of the Grad's moment method.
  In their paper
  \cite{TKO},
  TKO proved that
  the Eckart constraint (\ref{eq:PT-E}) in the particle frame
  cannot be compatible with the underlying relativistic Boltzmann equation
  for the first-order hydrodynamic equation.
  In spite of the first-order one,
  the TKO equation in the particle frame
  is free from the pathological properties
  \cite{Tsumura:2007wu}
  in contrast to the original Eckart equation
  with which the thermal equilibrium
  becomes unstable for a small perturbation
  \cite{hyd002}.

  One may naturally ask if the Eckart
  constraint (\ref{eq:PT-E})
  should be replaced with (\ref{eq:PT-TKO})
  even for the so-called second-order equation like Israel-Stewart one.
  And are any modifications needed to the constraints in the Landau frame?
  A purpose of this Letter
  \cite{QM09}
  is to answer these questions both by phenomenological and microscopic analyses.
  We shall see that
  the Eckart constraint should be replaced
  with the new one even in the second-order equation,
  while no modification is necessary
  for the constraints in the energy frame.
  We shall derive
  the second-order dissipative relativistic hydrodynamic equations
  in a generic frame with a continuous parameter $\theta$
  from the relativistic Boltzmann equation.
  We shall derive the relaxation terms
  for a generic frame with the new constraint,
  and present explicitly those in the energy and particle frames.
  We shall show that
  the viscosities are frame-independent
  but the relaxation times are generically frame-dependent
  in accordance with the observation by Betz et al.
  \cite{Betz:2008me},
  although the constraint to $\delta T^{\mu\nu}$ is quite different.
  
  \setcounter{equation}{0}
  \section{
    A general phenomenological derivation 
    of  relativistic dissipative hydrodynamic equations;\,
    existence of possible  extra terms in the dissipative terms
  }
  Let $T^{\mu\nu}$ and $N^\mu$ be
  the symmetric energy-momentum tensor and the particle-number vector
  of the system we consider, respectively.
  The total number of independent variables is fourteen,
  and the dynamical evolution of these variables
  are governed by the respective balance equations;
  \begin{eqnarray}
    \label{eq:2-001}
    \partial_\mu T^{\mu\nu} &=& 0,\\
    \label{eq:2-002}
    \partial_\mu N^\mu &=& 0.
  \end{eqnarray}
  
  With use of an arbitrary four vector $u^\mu$ with $u^\mu \, u_\mu = 1$,
  $T^{\mu\nu}$ and $N^\mu$ can be cast into the tensor-decomposed forms,
  \begin{eqnarray}
    \label{eq:2-003}
    T^{\mu\nu} &=& (e + \delta e) \, u^\mu
    \, u^\nu
    - (p + \delta p) \,
    \Delta^{\mu\nu} + q^\mu \, u^\nu + q^\nu \, u^\mu +
    \pi^{\mu\nu},\\
    \label{eq:2-004}
    N^\mu &=& (n + \delta n) \, u^\mu + \nu^\mu,
  \end{eqnarray}
  respectively.
  Here,
  $e+\delta e$, $p+\delta p$, and $n+\delta n$
  are the internal energy, pressure,
  and particle-number density in the dissipative system;
  $e + \delta e \equiv T_{ab} \, u^a \, u^b$,
  $p + \delta p \equiv -1/3 \, T_{ab} \, \Delta^{ab}$,
  and $n + \delta n \equiv N_a \, u^a$,
  with
  $e = e(T,\,\mu)$,
  $p = p(T,\,\mu)$,
  and $n = n(T,\,\mu)$
  being the corresponding quantities
  in the local equilibrium state
  characterized by the temperature $T$ and the chemical potential $\mu$.
  Note that we have made it explicit by
  $\delta e$, $\delta p$, and $\delta n$ that
  the dissipations may cause corrections
  to all these quantities,
  although only the correction to the pressure has been considered in the literature;
  $\delta p$ is identified with the bulk pressure $\Pi$.
  We emphasize that
  there is no persuading reasoning that
  only the pressure acquires corrections due to the dissipative process.
  The dissipative parts of the energy-momentum tensor and
  particle-number vector are identified as
  $\delta T^{\mu\nu} = \delta e \, u^\mu \, u^\nu + \delta p \, \Delta^{\mu\nu} +
  q^\mu \, u^\nu + q^\nu \, u^\mu + \pi^{\mu\nu}$ and
  $\delta N^{\mu} = \delta n \, u^\mu + \nu^\mu$, respectively.
  The energy flow relative to $u^\mu$ is denoted by $q^\mu$,
  $\nu^\mu$ is the flow of particle number relative to $u^\mu$,
  and finally
  $\pi^{\mu\nu}$ is the shear stress tensor;
  $q^\mu \equiv T_{ab} \, u^a \, \Delta^{b\mu}$,
  $\nu^\mu \equiv N_a \, \Delta^{a\mu}$,
  and $\pi^{\mu\nu} \equiv T_{ab} \, \Delta^{ab\mu\nu}$.
  Here
  the space-like, symmetric and traceless tensor
  $\Delta^{\mu\nu\rho\sigma} \equiv
  1/2 \, (\Delta^{\mu\rho} \, \Delta^{\nu\sigma} + \Delta^{\mu\sigma} \,
  \Delta^{\nu\rho} - 2/3 \, \Delta^{\mu\nu} \, \Delta^{\rho\sigma})$
  is introduced.
  One can easily confirm that
  $q^\mu \, u_\mu = 0$, $\nu^\mu \, u_\mu = 0$,
  $\pi^{\mu\nu} = \pi^{\nu\mu}$,
  and $u_\mu \, \pi^{\mu\nu} = \pi^\mu_{\,\,\,\mu} = 0$.
  This implies that
  the total number of independent components
  of $q^\mu$, $\nu^\mu$, and $\pi^{\mu\nu}$ is eleven.
  Since $T^{\mu\nu}$ and $N^\mu$
  have the fourteen components in total,
  $\delta e$, $\delta p$, and $\delta n$ have
  only one independent component other than $T$ and $\mu$.
  We take $\delta p= \Pi$ as the independent component as a natural choice,
  then $\delta e$ and $\delta n$ can be expressed as
  $\delta e = f_e \, \Pi$ and $\delta n = f_n \, \Pi$,
  where $f_e$ and $f_n$ are functions of $T$ and $\mu$;
  $f_e = f_e(T,\,\mu)$ and $f_n = f_n(T,\,\mu)$.
  Here we have assumed that the dissipative order of
  $\delta e$ and $\delta n$ are the same as
  that of $\delta p$ at most.
  We remark that
  although $f_e$ and $f_n$ may take finite values generically,
  the functional forms of
  $f_e$ and $f_n$ cannot be determined by the phenomenological theory,
  as those of  $e$, $p$, and $n$ can not, either.
  All the previous analyses assumed
  $f_e = f_n = 0$, which has not been recognized so far.
  
  Now we shall show that
  the just usual phenomenological derivation
  of the hydrodynamic equations
  in which the second law of thermodynamics is utilized
  allows the existence of $\delta e$ and $\delta n$, i.e.,
  finite values of $f_e$ and $f_n$,
  in the relativistic dissipative hydrodynamic equations.
  It is found that the essential point of the proof
  is the same for the first- and second-order equations
  where the entropy current $S^\mu$ is
  at most linear and bilinear with respect to
  $\Pi$, $q^\mu$, $\nu^\mu$, and $\pi^{\mu\nu}$, respectively,
  although the resulting mathematical expressions
  are much more complicated in the second-order one
  \cite{next002}.
  Thus we here take the first-order equation, for the sake of simplicity.
  The second-order equations with finite $f_e$ and $f_n$
  will be derived microscopically later in this article.
  So the entropy current is given by
  \begin{eqnarray}
    \label{eq:2-015}
    T \, S^\mu = p \, u^\mu + u_\nu \, T^{\mu\nu} -
    \mu \, N^\mu.
  \end{eqnarray}
  The second law of thermodynamics
  reads $\partial_\mu S^\mu \ge 0$.
  
  The divergence of $S^\mu$ is found to take the form
  \begin{eqnarray}
    \label{eq:2-016}
    \partial_\mu S^\mu = \Pi \, \Bigg[
      f_e\,D\frac{1}{T} - \frac{1}{T} \, \nabla^\mu u_\mu
      - f_n\,D\frac{\mu}{T}
      \Bigg]
    + q^\mu \, \Bigg[ \frac{1}{T} \, Du_\mu + \nabla_\mu
      \frac{1}{T} \Bigg] - \nu^\mu \,
    \nabla_\mu \frac{\mu}{T}
    + \pi^{\mu\nu} \, \frac{1}{T} \, \nabla_\mu u_\nu,
  \end{eqnarray}
  where
  $D \equiv u^a \, \partial_a$
  and
  $\nabla^\mu \equiv \Delta^{\mu a} \, \partial_a$.
  Here, we have used
  the conservation laws, Eq.'s (\ref{eq:2-001}) and (\ref{eq:2-002}),
  and the first law of thermodynamics,
  $D(p/T) + e \, D(1/T) - n \, D(\mu/T) = 0$.
  
  The frames define the flow velocity $u^{\mu}$ of the fluid:
  The flow velocity in the particle frame
  and the energy frame are
  defined by setting
  $u^\mu = N^\mu / \sqrt{N^\nu \, N_\nu}$
  and
  $u^\mu = T^{\mu a} \, u_a / \sqrt{T^{\nu b} \, u_b \, T_{\nu c} \, u^c}$, respectively
  \cite{mic001}.
  By these settings,
  a closed system of the relativistic dissipative
  hydrodynamic equations is obtained.
  Note that
  $u^\mu = N^\mu / \sqrt{N^\nu \, N_\nu}$
  ($u^\mu = T^{\mu a} \, u_a / \sqrt{T^{\nu b} \, u_b \, T_{\nu c} \, u^c}$)
  is equivalent to $\nu^\mu = 0$
  ($q^\mu = 0$).
  
  In the particle frame where $\nu^\mu = 0$,
  Eq.(\ref{eq:2-016}) is reduced to
  \begin{eqnarray}
  \label{eq:2-019}
  \partial_\mu S^\mu = \Pi \, \Bigg[
    f_e\,D\frac{1}{T} - \frac{1}{T} \, \nabla^\mu u_\mu
    - f_n\,D\frac{\mu}{T}
    \Bigg]
  + q^\mu \, \Bigg[ \frac{1}{T} \, Du_\mu + \nabla_\mu \frac{1}{T} \Bigg]
  + \pi^{\mu\nu} \, \frac{1}{T} \, \nabla_\mu u_\nu.
  \end{eqnarray}
  
  It is found that
  the following constitutive equations,
  \begin{eqnarray}
    \label{eq:2-021}
    \Pi &=& \zeta \, T \, \Bigg[
      f_e\,D \frac{1}{T} - \frac{1}{T} \, \nabla^\mu u_\mu
      - f_n\,D\frac{\mu}{T} \Bigg],\\
    \label{eq:2-022}
    q^\mu &=& - \lambda \, T^2 \, \Bigg[ \frac{1}{T} \, Du^\mu +
      \nabla^\mu \frac{1}{T} \Bigg],\\
    \label{eq:2-023}
    \pi^{\mu\nu} &=& 2 \, \eta \, \Delta^{\mu\nu\rho\sigma} \,
    \nabla_\rho u_\sigma,
  \end{eqnarray}
  guarantees the second law of thermodynamics,
  $\partial_\mu S^\mu \ge 0$,
  with $\zeta$, $\lambda$, and $\eta$
  being the bulk viscosity, heat conductivity, and shear viscosity, respectively.
  This is because the divergence $\partial_\mu S^\mu$ now becomes positive
  semi-definite;
  \begin{eqnarray}
    \label{eq:2-020}
    \partial_\mu S^\mu
    = \frac{\Pi^2}{\zeta T}
    - \frac{q^\mu q_\mu}{\lambda T^2}
    + \frac{\pi^{\mu\nu}\pi_{\mu\nu}}{2\eta T} \ge 0.
  \end{eqnarray}
  Thus we realize that
  there is nothing wrong
  with the resultant relativistic dissipative hydrodynamic equations
  with finite $f_e$ and $f_n$, or equivalently
  finite $\delta e$ and $\delta n$,
  which is compatible with the second law of thermodynamics.
  Eq.'s (\ref{eq:2-021})-(\ref{eq:2-023}) with
  a restricted condition $f_e = f_n = 0$
  are identical to the constitutive equations proposed by Eckart
  that are commonly used.
  
  In the energy frame where $q^\mu = 0$,
  we can obtain the constitutive equations
  in the same way as the particle-frame case with
  $f_e$ and $f_n$ being kept finite.
  The resultant equations are given by Eq.'s (\ref{eq:2-021}), (\ref{eq:2-023}), and
  \begin{eqnarray}
    \label{eq:2-024}
    \nu^\mu = \lambda \, \hat{h}^{-2} \,
    \nabla^\mu \frac{\mu}{T},
  \end{eqnarray}
  with $\hat{h} \equiv (e + p)/n\,T$ being the enthalpy.
  It is noted that
  these equations are reduced to
  the constitutive equations by Landau
  if we can set $f_e = f_n = 0$.
  
  By applying the above argument
  to the entropy current at most bilinear
  with respect to $\Pi$, $q^\mu$, $\nu^\mu$, and $\pi^{\mu\nu}$,
  we can obtain the relaxation equations
  with $f_e$ and $f_n$ being finite,
  which make up the so-called second-order
  relativistic dissipative hydrodynamic equations
  together with the conservation laws
  in Eq.'s (\ref{eq:2-001}) and (\ref{eq:2-002})
  \cite{next002}.
  
  Now the dissipative part of the energy-momentum tensor satisfies
  $u_{\mu} \, \delta T^{\mu\nu} \, u_{\nu} = \delta e = f_e \, \Pi$
  and
  $\delta T^{\mu}_{\,\,\,\mu} = \delta e - 3 \, \delta p = (f_e - 3) \, \Pi$.
  As emphasized before,
  the values of $f_e$ and $f_n$ can be determined only from a microscopic theory.
  The phenomenological theory cannot proceed further
  because no such logic to determine them is implemented in the theory.
  In the following section,
  we shall show that the microscopic theory
  gives $f_e = 3$ together with $f_n = 0$
  in the particle frame while
  $f_e = f_n = 0$ in the energy frame, and hence
  $\delta T^{\mu}_{\,\,\,\mu} = 0$
  but $u_{\mu} \, \delta T^{\mu\nu} \, u_{\nu} = 3 \, \Pi \neq 0$
  in the particle frame.
  This fact tells us that the usual constraint employed for the particle frame
  must be abandoned,
  and all the analyses based on this constraint should be redone.
  
  \setcounter{equation}{0}
  \section{
    Derivation of second-order equations
    as long wavelength and low frequency
    limit of relativistic Boltzmann equation
  }
  The argument so far is in the stage of thermodynamics
  where the argument is robust
  but the parameters such as $f_e$ and $f_n$
  as well as the equations of state $e$, $p$ and $n$ appearing in the theory
  remain undetermined.
  The problem which we encounter is how to reduce a dynamical equation
  to a slower one described with fewer dynamical variables.
  For this purpose,
  we will investigate the infrared limit of
  the relativistic Boltzmann equation
  with use of a powerful reduction method,
  the ``RG method''
  \cite{rgm001,env001,env004,HK02,env006}.
  
  The RG method is a systematic reduction theory of the dynamics
  leading to the coarse-graining of temporal and spatial scales.
  The full presentation of the reduction
  of the relativistic Boltzmann equation
  to the second-order hydrodynamic equation is technical and involved.
  So we here only present main results with key several equations,
  leaving the detailed account to another publication
  \cite{next002},
  although the derivation of a wide class of the first-order equations
  is presented
  in Ref.\cite{TKO}.
  
  We start with the simple relativistic Boltamann equation,
  \begin{eqnarray}
    \label{eq:2.1.1}
    p^\mu \, \partial_\mu \, f_p(x) = C[f]_p(x),
  \end{eqnarray}
  where $f_p(x)$ denotes the one-particle distribution function
  defined in the phase space $(x^{\mu} \,,\, p^{\mu})$
  with $p^\mu$ being the four momentum of the on-shell particle.
  The right-hand side of Eq.(\ref{eq:2.1.1}) is the collision integral,
  $C[f]_p(x) \equiv \frac{1}{2!} \, \sum_{p_1} \, \frac{1}{p_1^0} \,
  \sum_{p_2} \, \frac{1}{p_2^0} \, \sum_{p_3} \, \frac{1}{p_3^0} \, 
  \omega(p \,,\, p_1|p_2 \,,\, p_3) \, 
  ( f_{p_2}(x) \, f_{p_3}(x) - f_p(x) \, f_{p_1}(x) )$,
  where $\omega(p \,,\, p_1|p_2 \,,\, p_3)$ denotes
  the transition probability owing to the microscopic two-particle interaction.
  
  We are interested in the hydrodynamical regime
  where the time- and space-dependence of the physical quantities are small.
  In another word,
  the time and space entering the hydrodynamic equation are
  the ones coarse-grained from those in the kinetic equation.
  Thus we are lead to introduce a macroscopic Lorentz vector,
  $\Ren{a}^\mu_p(x)$
  which specifies the covariant coordinate system
  and we call the \textit{macroscopic-frame vector}.
  With use of $\Ren{a}^\mu_p(x)$,
  we define the macroscopic covariant coordinate system $(\tau \,,\, \sigma^\mu)$ as
  $\mathrm{d}\tau \equiv \Ren{a}_p^\mu(x) \, \mathrm{d}x_\mu$
  and
  $\varepsilon^{-1} \, \mathrm{d}\sigma^\mu
  \equiv ( g^{\mu\nu} - \Ren{a}_p^\mu(x)\Ren{a}_p^\nu(x)/\Ren{a}_p^2(x) )
  \, \mathrm{d}x_\nu
  \equiv \Ren{\Delta}_p^{\mu\nu}(x) \, \mathrm{d}x_\nu$.
  We note that the small quantity $\varepsilon$ has been introduced to tag
  that the space derivatives are small for the system we are interested in.
  $\varepsilon$ may be identified with the ratio of the average
  particle distance over the mean free path,
  i.e., the Knudsen number.
  
  In this coordinate system, Eq.(\ref{eq:2.1.1}) can be cast into
  \begin{eqnarray}
    \label{eq:2.1.6}
    \frac{\partial}{\partial \tau} f_p(\tau \,,\, \sigma)
    = \frac{1}{p \cdot \Ren{a}_p(\tau \,,\, \sigma)} \, C[f]_p(\tau \,,\, \sigma)
    - \varepsilon \, \frac{1}{p \cdot \Ren{a}_p(\tau \,,\, \sigma)} \,
    p \cdot \Ren{\nabla} f_p(\tau \,,\, \sigma),
  \end{eqnarray}
  where $\Ren{a}_p^\mu(\tau \,,\, \sigma) \equiv \Ren{a}_p^\mu(x)$,
  $\Ren{\Delta}_p^{\mu\nu}(\tau \,,\, \sigma) \equiv \Ren{\Delta}_p^{\mu\nu}(x)$,
  and $f_p(\tau \,,\, \sigma) \equiv f_p(x)$.
  Since $\varepsilon$ appears in front of
  $\Ren{\nabla}^\mu \equiv \Ren{\Delta}_p^{\mu\nu}(\tau \,,\, \sigma)
  \, \frac{\partial}{\partial \sigma^\nu}$,
  Eq.(\ref{eq:2.1.6}) has a form to which the perturbative expansion
  with respect to $\varepsilon$ can be applied.
  In the perturbative expansion, we shall take the coordinate system
  where $\Ren{a}_p^\mu(\tau \,,\, \sigma)$ has no $\tau$ dependence,
  i.e., $\Ren{a}_p^\mu(\tau \,,\, \sigma) = \Ren{a}_p^\mu(\sigma)$.
  
  The zeroth-order approximate solution we construct
  is a stationary solution,
  which is identical to a local equilibrium distribution function
  given by the Juetner function
  $f^{\mathrm{eq}}_p \equiv  (2\,\pi)^{-3} \, \exp [(\mu - p \cdot u)/T]$.
  Note that this solution
  contains five would-be integration constants,
  $T$, $\mu$, and $u^\mu$ with $u^\mu \, u_\mu = 1$,
  which can be identified with
  the temperature, the chemical potential,
  and the fluid velocity, respectively.
  
  The collision integral is expanded around the zeroth-order solution
  and is reduced to the linear operator
  $A_{pq} \equiv (p \cdot \Ren{a}_p)^{-1} \, \frac{\partial}{\partial
  f_q} C[f^\mathrm{eq}]_p$.
  Furthermore, it is found to be convenient to
  convert $A_{pq}$ to
  $L_{pq} \equiv f^{\mathrm{eq}-1}_p \, A_{pq} \, f^{\mathrm{eq}}_q =
  [ f^{\mathrm{eq}-1} \, A \, f^{\mathrm{eq}} ]_{pq}$,
  with the diagonal matrix
  $f^\mathrm{eq}_{pq} \equiv f^\mathrm{eq}_p \delta_{pq}$.
  We also define the inner product between arbitrary vectors
  $\varphi_p$ and $\psi_p$ by
  \begin{eqnarray}
    \label{eq:3.2.11}
    \langle \, \varphi \,,\, \psi \,
    \rangle \equiv \sum_{p} \, \frac{1}{p^0} \, (p \cdot \Ren{a}_p) \,
    f^{\mathrm{eq}}_p \, \varphi_p \, \psi_p.
  \end{eqnarray}
  With this inner product, we can define a normed linear space.
  
  Now the first-order solution is given
  in terms of the five zero modes of $L$,
  $\varphi^\alpha_{0p} = (1,\,p^\mu)$.
  The corresponding variables are
  just $T$, $\mu$, and $u^\mu$ with $u_{\mu}u^{\mu} = 1$.
  The zero modes span a linear space $\mathrm{P}_0$,
  which is an invariant manifold for the asymptotic dynamics
  of the relativistic Boltamann equation
  in the terminology in the dynamical systems
  \cite{env004,inv-manifold}.
  
  Then the second-order solution is given by incorporating
  the next slow modes, which span a linear space $\mathrm{P}_1$.
  We naturally require $\mathrm{P}_1$ is orthogonal to
  $\mathrm{P}_0 $, that is,  $\mathrm{P}_0 \perp \mathrm{P}_1$.
  We find that $\mathrm{P}_1$ is expanded by the bilinear forms of momenta;
  $\varphi^{\mu\nu}_{1p} \equiv [ Q_0 \, \tilde{\varphi}^{\mu\nu} ]_p$,
  where $\tilde{\varphi}^{\mu\nu}_p \equiv p^\mu \, p^\nu$, and
  $Q_0$ is the projection to complement to $\mathrm{P}_0$.
  By definition,
  $\langle \, \varphi^{\mu\nu}_1 \,,\, \varphi^\alpha_0 \, \rangle =
  0$ is satisfied.
  Note that the dimension of  $\varphi^{\mu\nu}_{1p}$ is nine,
  which correspond to
  the number of the new would-be integration constants,
  $\Pi$, $J^\mu$ with $J^\mu \, u_\mu = 0$,
  and $\pi^{\mu\nu}$ with
  $\pi^{\mu\nu} = \pi^{\nu\mu}$
  and
  $\pi^{\mu\nu} \, u_\nu = \pi^\mu_{\,\,\,\mu} = 0$.
  
  A generic choice of the macroscopic frame vector is
  $\Ren{a}^\mu_p = ((p\cdot u) \, \cos \theta + m \,
  \sin\theta)/(p\cdot u) \, u^\mu$,
  where $\theta$ is a parameter defining the frame.
  For example, $\theta = 0$ ($\theta = \pi/2$) gives
  the energy (particle) frame.
  
  The resultant generic relaxation equations
  of the second-order hydrodynamic equation with
  $\theta$ being kept are
  \begin{eqnarray}
    \Pi &=& X_\Pi - \tau_\Pi \, D\Pi - \ell_{\Pi J} \, \nabla^a J_a
    + X_{\Pi\Pi} \, \Pi + X_{\Pi J}^a \, J_a + X_{\Pi\pi}^{ab} \,
    \pi_{ab},\\
    J^\mu &=& X^\mu_J - \tau_J \, \Delta^{\mu a} \, DJ_a - \ell_{J\Pi}
    \, \nabla^\mu \Pi - \ell_{J\pi} \, \Delta^{\mu abc} \, \nabla_a
    \pi_{bc} + X_{J\Pi}^\mu \, \Pi + X_{JJ}^{\mu a} \, J_a +
    X_{J\pi}^{\mu ab} \, \pi_{ab},\\
    \pi^{\mu\nu} &=& X^{\mu\nu}_\pi - \tau_\pi \, \Delta^{\mu\nu ab}
    \, D \pi_{ab} - \ell_{\pi J} \, \Delta^{\mu\nu ab} \, \nabla_a J_b
    + X_{\pi\Pi}^{\mu\nu} \, \Pi + X_{\pi J}^{\mu\nu a} \, J_a +
    X_{\pi\pi}^{\mu\nu ab} \, \pi_{ab}.
  \end{eqnarray}
  Here, $X_\Pi$, $X^\mu_J$, and $X^{\mu\nu}_\pi$
  are the thermodynamic forces;
  their simple forms retaining only
  $X_\Pi$, $X^\mu_J$, and $X^{\mu\nu}_\pi$
  are the usual constitutive equations.
  The relaxation equations of
  $\Pi$, $J^\mu$, and $\pi^{\mu\nu}$ are characterized by
  the relaxation times  $\tau_\Pi$, $\tau_J$, and $\tau_\pi$, 
  while  $\ell_{\Pi J}$, $\ell_{J \Pi}$, $\ell_{J \pi}$, and $\ell_{\pi J}$
  mean the relaxation lengths.
  The correction to the thermodynamic forces
  $X_\Pi$, $X^\mu_J$, and $X^{\mu\nu}_\pi$ are given by
  $X_{\Pi\Pi}$, $X_{\Pi J}^a$, $X_{\Pi \pi}^{ab}$,
  $X_{J \Pi}^\mu$, $X_{J J}^{\mu a}$, $X_{J \pi}^{\mu ab}$,
  $X_{\pi\Pi}^{\mu\nu}$, $X_{\pi J}^{\mu\nu a}$, and $X_{\pi \pi}^{\mu\nu ab}$.
  
  The continuity equations of the second-order equation
  in the energy frame
  is found to be given by setting $\theta = 0$
  as in the first-order case
  \cite{TKO}
  and read $\partial_\mu T^{\mu\nu} = 0$
  and
  $\partial_\mu N^\mu = 0$,
  where
  \begin{eqnarray}
    T^{\mu\nu} &=& e \, u^\mu \, u^\nu - (p + \Pi) \, \Delta^{\mu\nu}
    + \pi^{\mu\nu},\\
    N^\mu &=& n \, u^\mu + J^\mu.
  \end{eqnarray}
  The thermodynamic forces are
  $X_\Pi = - \zeta \, \nabla^a u_a$,
  $X^\mu_J = \lambda \,  \hat{h}^{-2} \, \nabla^\mu (\mu/T)$,
  and
  $X^{\mu\nu}_\pi = 2 \, \eta \, \Delta^{\mu\nu ab} \, \nabla_a u_b$,
  which clearly show that $f_n = 0$ and $f_e=0$ as was anticipated.
  
  The energy-momentum tensor and particle-number vector
  in the particle frame with $\theta = \pi/2$ read
  \begin{eqnarray}
    T^{\mu\nu} &=& (e + 3 \, \Pi) \, u^\mu \, u^\nu - (p + \Pi) \, \Delta^{\mu\nu}
    + u^\mu \, J^\nu + u^\nu \, J^\mu + \pi^{\mu\nu},\\
    N^\mu &=& n \, u^\mu,
  \end{eqnarray}
  and
  $X_\Pi = - \zeta \, (3 \, \gamma - 4 )^{-2} \,
  (\nabla^a u_a - 3 \, T \, DT^{-1})$,
  $X^\mu_J = - \lambda \,  T^2 \, (\nabla^\mu T^{-1} + T^{-1} \, D u^\mu )$,
  and
  $X^{\mu\nu}_\pi = 2 \, \eta \, \Delta^{\mu\nu ab} \, \nabla_a u_b$,
  where
  $\gamma \equiv 1 + (z^2 - \hat{h}^2 + 5\,\hat{h} - 1)^{-1}$ is the ratio
  of the specific heats.
  Thus we find that $f_e = 3$ with $f_n = 0$, as we announced.
  
  Although we have obtained the relaxation equations
  for the dissipative forces $\Pi$, $J^{\mu}$, and $\pi ^{\mu \nu}$
  for arbitrary $\theta$
  \cite{next002},
  we shall only write down them for two typical frames,
  i.e., the energy ($\theta=0$)
  and the particle ($\theta=\pi/2$) frames for the sake of the space.
  
  (A) In the energy frame ($\theta=0$):
  \begin{eqnarray}
    \label{eq:EF_Pi}
    \Pi &=& - \zeta \, \nabla^a u_a - \tau_\Pi \, D\Pi - \ell_{\Pi J}
    \, \nabla^a J_a\nonumber\\
    &&{}- \frac{1}{2} \, \tau_\Pi \, \Bigg\{
    \kappa_\Pi \, \nabla^a u_a
    + \frac{\zeta \, T}{\tau_\Pi} \,
    \partial_a \Big(\frac{\tau_\Pi}{\zeta \, T} \, u^a\Big)
    \Bigg\} \,
    \Pi\nonumber\\
    &&{}- \frac{1}{2} \, \ell_{\Pi J} \, \Bigg\{
    \kappa^{(0)}_{\Pi J} \, \nabla^a \frac{\mu}{T}
    -
    \kappa^{(1)}_{\Pi J} \, D u^a
    + \frac{\zeta \, T}{\ell_{\Pi J}} \,
    \partial_b \Big(\frac{\ell_{\Pi J}}{\zeta\,T} \,
    \Delta^{bc}\Big)\,\Delta_c^{\,\,\,a}
    \Bigg\} \, J_a\nonumber\\
    &&{}-\frac{1}{2} \, \ell_{\Pi\pi} \, \Bigg\{
    - \kappa_{\Pi\pi} \, \Delta^{abcd} \, \nabla_c u_d
    \Bigg\} \, \pi_{ab},\\
    \label{eq:EF_J}
    J^\mu &=& \lambda \, \hat{h}^{-2} \, \nabla^\mu
    \frac{\mu}{T} - \ell_{J\Pi}
    \, \nabla^\mu \Pi
    - \tau_J \, \Delta^{\mu a} \, DJ_a
    - \ell_{J\pi} \, \Delta^{\mu abc} \, \nabla_a
    \pi_{bc}\nonumber\\
    &&{}- \frac{1}{2} \, \ell_{J\Pi} \, \Bigg\{
    \kappa_{J\Pi}^{(0)} \, \nabla^\mu \frac{\mu}{T}
    -
    \kappa_{J\Pi}^{(1)} \, Du^\mu
    + \frac{\lambda \, \hat{h}^{-2}}{\ell_{J\Pi}} \,
    \Delta^\mu_{\,\,\,a} \,
    \partial_b \Big(
    \frac{\ell_{J\Pi}}{\lambda \, \hat{h}^{-2}} \, \Delta^{ab}
    \Big)
    \Bigg\} \,
    \Pi\nonumber\\
    &&{}- \frac{1}{2} \, \tau_J \, \Bigg\{
    \Delta^{\mu a} \, \Bigg[
      \kappa_J^{(0)} \, \nabla^b u_b
      + \frac{\lambda \, \hat{h}^{-2}}{\tau_J} \,
      \partial_b \Big(\frac{\tau_J}{\lambda \, \hat{h}^{-2}} \, u^b\Big)
      \Bigg]
    - 2 \, \kappa_J^{(1)} \, \Delta^{\mu abc} \, \nabla_b u_c
    - 2 \, \omega^{\mu a}
    \Bigg\} \, J_a\nonumber\\
    &&{}- \frac{1}{2} \, \ell_{J\pi} \, \Bigg\{
    \Delta^{\mu cab} \, \Big(
      \kappa^{(0)}_{J\pi} \, \nabla_c \frac{\mu}{T}
      -
      \kappa^{(1)}_{J\pi} \, D u_c
      \Big)
    + \frac{\lambda\,\hat{h}^{-2}}{\ell_{J\pi}} \, \Delta^\mu_{\,\,\,c} \,
    \partial_d \Big(\frac{\ell_{J\pi}}{\lambda\,\hat{h}^{-2}} \, \Delta^{cdef}\Big) \,
    \Delta_{ef}^{\,\,\,\,\,\,ab}
    \Bigg\} \, \pi_{ab},\\
    \label{eq:EF_pi}
    \pi^{\mu\nu} &=& 2 \, \eta \, \Delta^{\mu\nu ab} \, \nabla_a
    u_b - \ell_{\pi J} \, \Delta^{\mu\nu ab} \, \nabla_a J_b
    - \tau_\pi \, \Delta^{\mu\nu ab}
    \, D \pi_{ab}\nonumber\\
    &&{}- \frac{1}{2} \, \ell_{\pi\Pi} \, \Bigg\{
    - \kappa_{\pi\Pi} \, \Delta^{\mu\nu ab} \,
    \nabla_a u_b
    \Bigg\} \, \Pi\nonumber\\
    &&{}- \frac{1}{2} \, \ell_{\pi J} \, \Bigg\{
    \Delta^{\mu\nu ba} \, \Big(
      \kappa^{(0)}_{\pi J} \, \nabla_b \frac{\mu}{T}
      -
      \kappa^{(1)}_{\pi J} \, D u_b
      \Big)
      + \frac{\eta\,T}{\ell_{\pi J}} \,
    \Delta^{\mu\nu}_{\,\,\,\,\,\,bc} \, \partial_d
    \Big(\frac{\ell_{\pi J}}{\eta\,T} \,
    \Delta^{bcde}\Big) \, \Delta_e^{\,\,\,a}
    \Bigg\} \, J_a\nonumber\\
    &&{}- \frac{1}{2} \, \tau_\pi \, \Bigg\{
    \Delta^{\mu\nu ab} \,
    \Bigg[
      \kappa^{(0)}_{\pi} \, \nabla^c u_c
      + \frac{\eta\,T}{\tau_\pi} \,
      \partial_c \Big(\frac{\tau_\pi}{\eta\,T} \, u^c\Big)
      \Bigg]
    - 4 \, \kappa^{(1)}_\pi \, \Delta^{\mu\nu ce} \,
    \Delta_e^{\,\,\,dab} \, \Delta_{cd}^{\,\,\,\,\,\,fg} \,
    \nabla_f u_g
    - 4 \, \Delta^{\mu\nu ce} \, \Delta_{e}^{\,\,\,dab}
    \, \omega_{cd}
    \Bigg\} \, \pi_{ab},\nonumber\\
  \end{eqnarray}
  where $\omega^{\mu\nu} \equiv (\nabla^\mu u^\nu - \nabla^\nu
  u^\mu)/2$ is the vorticity.

  (B) In the particle frame ($\theta=\pi/2$):
  \begin{eqnarray}
    \label{eq:PF_Pi}
    \Pi &=& - \zeta \, (3 \, \gamma - 4 )^{-2} \,
    \Big(\nabla^a u_a - 3 \, T \, D\frac{1}{T}\Big) - \tau_\Pi \, D\Pi - \ell_{\Pi J}
    \, \nabla^a J_a\nonumber\\
    &&{}- \frac{1}{2} \, \tau_\Pi \, \Bigg\{
    \kappa_\Pi \, \nabla^a u_a
    + \frac{\zeta \, (3 \, \gamma - 4 )^{-2} \, T}{\tau_\Pi} \,
    \partial_a \Big(\frac{\tau_\Pi}{\zeta \, (3 \, \gamma - 4 )^{-2} \, T} \, u^a\Big)
    \Bigg\} \,
    \Pi\nonumber\\
    &&{}- \frac{1}{2} \, \ell_{\Pi J} \, \Bigg\{
    \kappa^{(0)}_{\Pi J} \, \nabla^a \frac{\mu}{T}
    -
    \kappa^{(1)}_{\Pi J} \, D u^a
    + \frac{\zeta \, (3 \, \gamma - 4 )^{-2} \, T}{\ell_{\Pi J}} \,
    \partial_b \Big(\frac{\ell_{\Pi J}}{\zeta\,(3 \, \gamma - 4 )^{-2} \, T} \,
    \Delta^{bc}\Big)\,\Delta_c^{\,\,\,a}
    \Bigg\} \, J_a\nonumber\\
    &&{}-\frac{1}{2} \, \ell_{\Pi\pi} \, \Bigg\{
    - \kappa_{\Pi\pi} \, \Delta^{abcd} \, \nabla_c u_d
    \Bigg\} \, \pi_{ab},\\
    \label{eq:PF_J}
    J^\mu &=&
    - \lambda \,  T^2 \, \Big(\nabla^\mu
    \frac{1}{T} + \frac{1}{T} \, D u^\mu \Big) - \ell_{J\Pi}
    \, \nabla^\mu \Pi
    - \tau_J \, \Delta^{\mu a} \, DJ_a
    - \ell_{J\pi} \, \Delta^{\mu abc} \, \nabla_a
    \pi_{bc}\nonumber\\
    &&{}- \frac{1}{2} \, \ell_{J\Pi} \, \Bigg\{
    \kappa_{J\Pi}^{(0)} \, \nabla^\mu \frac{\mu}{T}
    -
    \kappa_{J\Pi}^{(1)} \, Du^\mu
    + \frac{\lambda \, T^{2}}{\ell_{J\Pi}} \,
    \Delta^\mu_{\,\,\,a} \,
    \partial_b \Big(
    \frac{\ell_{J\Pi}}{\lambda \, T^{2}} \, \Delta^{ab}
    \Big)
    \Bigg\} \,
    \Pi\nonumber\\
    &&{}- \frac{1}{2} \, \tau_J \, \Bigg\{
    \Delta^{\mu a} \, \Bigg[
      \kappa_J^{(0)} \, \nabla^b u_b
      + \frac{\lambda \, T^{2}}{\tau_J} \,
      \partial_b \Big(\frac{\tau_J}{\lambda \, T^{2}} \, u^b\Big)
      \Bigg]
    - 2 \, \kappa_J^{(1)} \, \Delta^{\mu abc} \, \nabla_b u_c
    - 2 \, \omega^{\mu a}
    \Bigg\} \, J_a\nonumber\\
    &&{}- \frac{1}{2} \, \ell_{J\pi} \, \Bigg\{
    \Delta^{\mu cab} \, \Big(
      \kappa^{(0)}_{J\pi} \, \nabla_c \frac{\mu}{T}
      -
      \kappa^{(1)}_{J\pi} \, D u_c
      \Big)
      + \frac{\lambda\,T^{2}}{\ell_{J\pi}} \, \Delta^\mu_{\,\,\,c} \,
      \partial_d \Big(\frac{\ell_{J\pi}}{\lambda\,T^{2}} \, \Delta^{cdef}\Big) \,
    \Delta_{ef}^{\,\,\,\,\,\,ab}
    \Bigg\} \, \pi_{ab},
  \end{eqnarray}
  and Eq.(\ref{eq:EF_pi}).
  Note that the effective bulk viscosity $\zeta_\mathrm{eff} \equiv
  \zeta \, (3 \, \gamma - 4 )^{-2}$ \cite{Tsumura:2007wu}
  appears in Eq.(\ref{eq:PF_Pi}).

  Here we have introduced
  the new coefficients,
  $\ell_{\Pi\pi}$, $\ell_{\pi\Pi}$,
  $\kappa_\Pi$,
  $\kappa^{(0)}_{\Pi J}$,
  $\kappa^{(1)}_{\Pi J}$,
  $\kappa_{\Pi\pi}$,
  $\kappa^{(0)}_{J \Pi}$,
  $\kappa^{(1)}_{J \Pi}$,
  $\kappa^{(0)}_J$,
  $\kappa^{(1)}_J$,
  $\kappa^{(0)}_{J\pi}$,
  $\kappa^{(1)}_{J\pi}$,
  $\kappa_{\pi\Pi}$,
  $\kappa^{(0)}_{\pi J}$,
  $\kappa^{(1)}_{\pi J}$,
  $\kappa^{(0)}_\pi$,
  and
  $\kappa^{(1)}_\pi$,
  which are complicated functions of $T$ and $\mu$ \cite{next002}.

\begin{figure}
    \begin{center}
      \begin{minipage}{.45\linewidth}
        \includegraphics[width=\linewidth]{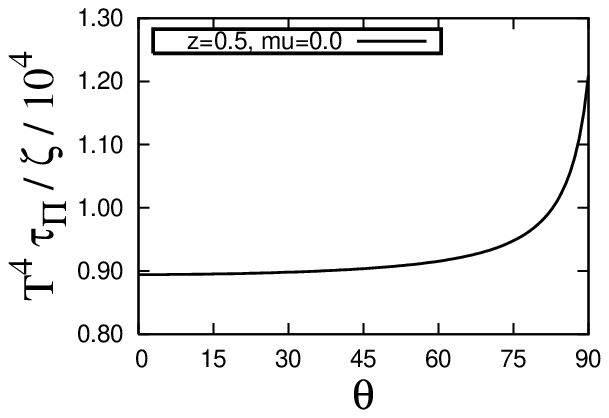}
      \end{minipage}
      \begin{minipage}{.45\linewidth}
        \includegraphics[width=\linewidth]{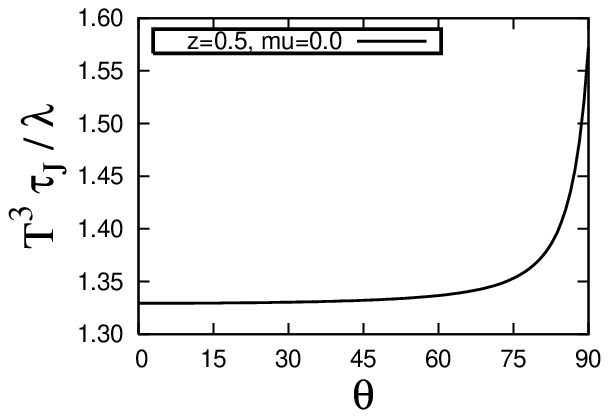}
      \end{minipage}
    \end{center}
    \caption{
      \label{fig:1}
      The $\theta$ dependence of $\tau_\Pi$ and $\tau_J$
      at $m/T = 0.5$ and $\mu/T = 0.0$.
      We normalized the relaxation times
      by the corresponding transport coefficients.
      The energy and particle frames correspond to
      $\theta=0$ and $\pi/2$.
    }
  \end{figure}

  We have found that the relaxation times are frame dependent except for $\tau_\pi$
  while the transport coefficients
  such as the viscosities and the thermal conductivity are frame independent.
  For a demonstration of the frame-dependence of the relaxation times,
  we show in FIG.\ref{fig:1} the frame ($\theta$) dependence of
  $\tau_\Pi$ and $\tau_J$,
  which tends to increase when the frame changes
  from the energy to particle frame.

  \section{
    Brief summary
  }
  In summary, we have derived
  the second-order dissipative relativistic hydrodynamic equations
  in a generic frame with a continuous parameter $\theta$;
  the generic frame is reduced
  to the energy and particle frame with the parameter choice
  $\theta=0$ and $\pi/2$, respectively.
  A notable point of our result is that
  the dissipative part of the symmetric energy-momentum tensor
  $\delta T^{\mu\nu}$ in the \textit{particle} frame
  satisfies the equality
  $\delta T^{\mu}_{\,\,\,\mu} = 0$,
  in contrast to the usual choice
  $u_{\mu} \, \delta T^{\mu\nu} \, u_{\nu} = 0$,
  while $\delta T^{\mu\nu}$ of our derived equation
  in the \textit{energy} frame
  satisfies the usual constraint $u_{\mu} \, \delta T^{\mu\nu} \, u_\nu = 0$.
  We emphasize that
  this novel equality in the particle frame
  is a consequence of the derivation
  based on the renormalization-group method,
  a powerful method for the reduction of dynamical systems.
  We note that
  the same constraints were also derived
  for the first-order dissipative relativistic hydrodynamic equation
  \cite{TKO,Tsumura:2007wu}.
  We have also shown that
  the phenomenological derivation
  based on the second law of thermodynamics allows that
  $u_{\mu} \, \delta T^{\mu\nu} \, u_{\nu}$ can be proportional
  to the bulk pressure $\Pi$
  and non-vanishing in the particle frame.
  Indeed, our microscopic derivation shows that
  $u_{\mu} \, \delta T^{\mu\nu} \, u_{\nu} = 3 \, \Pi$.
  We have presented the relaxation equations
  in the energy and particle frames,
  explicitly as typical examples,
  although we have obtained the microscopic expressions
  for them in a more generic frame
  \cite{next002}.
  We have shown that
  the viscosities are frame-independent
  but the relaxation times are generically frame-dependent,
  as depicted in FIG.\ref{fig:1}.
  The detailed derivation of the equations
  and discussions on the phenomenological consequences
  of the hydrodynamic equations thus obtained
  will be discussed in forthcoming papers
  \cite{next002}.
  
  \section*{acknowledgment}
  We thank Tetsu Hirano for his interest in this work and discussions.
  T.K. thanks Dirk Rischke for his interest in our work.
  This work was partially supported
  by a Grant-in-Aid for Scientific Research by the Ministry of Education,
  Culture, Sports, Science and Technology (MEXT) of Japan (No.20540265),
  by Yukawa International Program for Quark-Hadron Sciences,
  and by the Grant-in-Aid for the global COE program
  ``The Next Generation of Physics, Spun from Universality and Emergence''
  from MEXT.


\begin{thebibliography}{40}
  \bibitem{qcd001}
    See review articles,
    P.Huovinen,
    in ``Quark Gluon Plasma 3'', ed. R.C.Hwa and X.N.Wang,
    (World Scientific, Singapore), p.600;\,
    P.F.Kolb and U.W.Heinz, ibid, p.634.
    
  \bibitem{qcd002}
    T. Hirano and K. Tsuda,
    {Phys. Rev. C} \textbf{66}, 054905 (2002);
    D. Teaney,
    {Phys. Rev. C} \textbf{68}, 034913 (2003).
    
  \bibitem{qcd003}
    M.Gyulassy and L.McLerran,
    {Nucl.Phys.A} \textbf{750}, 30 (2005).
    
  \bibitem{Hirano:2008hy}
    T. Hirano, N. van der Kolk and A. Bilandzic,
    arXiv:0808.2684 [nucl-th].

  \bibitem{Romatschke:2009im}
    P. Romatschke,
    arXiv:0902.3663 [hep-ph].
    
  \bibitem{TKO}
    K. Tsumura, T. Kunihiro and K. Ohnishi,
    {Phys. Lett. B} \textbf{646}, 134 (2007).
    
  \bibitem{rgm001}
    L. Y. Chen, N. Goldenfeld and Y. Oono,
    {Phys. Rev. Lett.} \textbf{73}, 1311 (1994);
    {Phys. Rev. E} \textbf{54}, 376 (1996).
    
  \bibitem{env001}
    T. Kunihiro,
    {Prog. Theor. Phys.} \textbf{94}, 503 (1995); \textbf{95}, 835 (1996) (E);
    {Jpn. J. Ind. Appl. Math.} \textbf{14}, 51 (1997);
    {Prog. Theor. Phys.} \textbf{97}, 179 (1997).
    
  \bibitem{env004}
    S.-I. Ei, K. Fujii, and T. Kunihiro,
    {Ann. Phys.} \textbf{280}, 236 (2000).
    
  \bibitem{HK02}
    Y. Hatta and T. Kunihiro,
    {Ann. Phys.} \textbf{298}, 24 (2002).
    
  \bibitem{env006}
    T. Kunihiro and K. Tsumura,
    {J. Phys. A} \textbf{39}, 8089 (2006).
    
  \bibitem{hen002}
    L. D. Landau and E. M. Lifshitz,
    \textit{Fluid Mechanics}, (Pergamon Press, London, 1959).
    
  \bibitem{hen001}
    C. Eckart,
    {Phys. Rev.} \textbf{58}, 919 (1940).
    
  \bibitem{hen003}
    W. Israel,
    {Ann. Phys.} \textbf{100}, 310 (1976).
    
  \bibitem{mic004}
    W. Israel and J. M. Stewart,
    {Ann. Phys.} \textbf{118}, 341 (1979).
    
  \bibitem{mic001}
    S. R. de Groot, W. A. van Leeuwen and Ch. G. van Weert,
    \textit{Relativistic Kinetic Theory}, (Elsevier North-Holland, 1980).
    
  \bibitem{Betz:2008me}
    B. Betz, D. Henkel and D. H. Rischke,
    arXiv:0812.1440 [nucl-th].
    
  \bibitem{grad}
    H. Grad,
    {Comm. Pure Appl. Math.} \textbf{2}, 331 (1949).
    
  \bibitem{marle:69}
    C. Marle,
    Annales de l'institut Henri Poincare (A) Physique theorique,
    \textbf{10}, 67 (1969); 127 (1969).
    Marle's theory is reviewed in \cite{mic003,cercignani02}.
    
  \bibitem{mic003}
    J. M. Stewart,
    \textit{Non-Equilibrium Relativistic Kinetic Theory}
    (Lecture Notes in Physics No. 10; Springer, Berlin, 1971).
    
  \bibitem{Tsumura:2007wu}
    K. Tsumura and T. Kunihiro,
    {Phys. Lett. B} \textbf{668}, 425 (2008).
    
  \bibitem{hyd002}
    W. A. Hiscock and L. Lindblom,
    {Phys. Rev. D} \textbf{31}, 725 (1985).
    
  \bibitem{QM09}
    K. Tsumura and T. Kunihiro;
    talk presented at JPS meeting at Yamagata University, September 20, 2008;
    T. Kunihiro, Y. Minami and K. Tsumura;
    talk presented at Quark Matter 2009, Knoxville, March, 2009.
    
  \bibitem{next002}
    K. Tsumura and T. Kunihiro,
    in preparation.
    
  \bibitem{inv-manifold}
    See for example,
    J. Guckenheimer and P. Holmes,
    ``Nonlinear Oscillators, Dynamical Systems, and Bifurcations of Vector Fields''
    Springer-Verlag, 1983.
    
  \bibitem{cercignani02}
    C. Cercignani and G. M. Kremer,
    \textit{The relativistic Boltzmann equation: theory and applications},
    (Birkhauser, 2002);
    Progress in mathematical physics v. \textbf{22}.
  \end{thebibliography}
\end{document}